\documentclass[12pt,a4]{article}
\usepackage{amsmath}
\usepackage{bm}
\usepackage{amsfonts}
\usepackage{amssymb}
\usepackage{graphics}
\usepackage[normal]{caption2}
\usepackage{subfigure}
\usepackage{rotating}
\usepackage{citesort}
\def\be{\begin{equation}}
\def\ee{\end{equation}}
\def\ba{\begin{array}}
\def\ea{\end{array}}
\def\bea{\begin{eqnarray}}
\def\eea{\end{eqnarray}}

\begin{document}
\baselineskip 20pt \setlength\tabcolsep{2.5mm}
\renewcommand\arraystretch{1.5}
\setlength{\abovecaptionskip}{0.1cm}
\setlength{\belowcaptionskip}{0.5cm}
\begin{center} {\large\bf On the participant-spectator matter and thermalization of neutron-rich systems in heavy-ion collisions}\\
\vspace*{0.4cm}
{\bf Sakshi Gautam and Rajeev K. Puri}\footnote{Email:~rkpuri@pu.ac.in}\\
{\it  Department of Physics, Panjab University, Chandigarh -160
014, India.\\}
\end{center}
We study the participant-spectator matter at the energy of
vanishing flow for neutron-rich systems. Our study reveals similar
behaviour of participant-spectator for neutron-rich systems as for
stable systems and also points towards nearly mass independence
behaviour of participant-spectator matter for neutron-rich systems
at the energy of vanishing flow. We also study the thermalization
reached in the reactions of neutron-rich systems.


\newpage
\baselineskip 20pt

The isospin physics has attracted lot of attention of the present
nuclear physics researchers around the world for the past decade.
The establishment and upcoming radioactive ion beam (RIB)
facilities provide a major boon to this \cite{rib1,rib2}. RIBs
provide the possibility to study the nuclear matter under the
extreme conditions of isospin asymmetry. Heavy-ion reactions
induced by neutron-rich matter provide a unique opportunity to
explore the isospin dependence of in-medium nuclear interactions,
since isospin degree of freedom plays an important role in
heavy-ion collisions through both nuclear equation of state (EOS)
and in-medium nucleon-nucleon cross section. Role of isospin
degree of freedom has been investigated in collective flow and its
disappearance (energy at which flow disappears is called energy of
vanishing flow (EVF) or balance energy (E$_{bal}$)) for the past
decade. Isospin effects in flow were first predicted by Pak
\emph{et al}. \cite{pak} These isospin effects are due to the
competition among various reaction mechanisms like Coulomb force,
symmetry energy, isospin dependence of cross section, and surface
effects. Though many studies are available on the energy of
vanishing flow, very few studies exist in the literature that are
carried out to study other heavy-ion phenomena at the balance
energy \cite{he,bali,soff,sood1}. Balance energy results due to
the counterbalancing of the attractive mean field and repulsive
nucleon-nucleon collisions. In terms of theoretical description,
it is the relative dominance of real and imaginary parts of
G-matrix which decides the fate of a reaction. The dominance of
nucleon-nucleon collisions at high incident energies makes the
imaginary part very significant. However, both real and imaginary
parts of complex G-matrix are equally important at intermediate
energies. This picture can also be looked in terms of
participant-spectator matter and fireball concept. In Ref.
\cite{sood1}, Sood and Puri studied the participant-spectator
matter and nuclear dynamics for stable systems. The study revealed
that for stable systems (N/Z $\simeq$ 1), participant-spectator
matter at E$_{bal}$ is quite insensitive to the mass of the
colliding system. It, therefore, can act as a barometer for the
study of balance energy. No study exists in literature to
demonstrate how these observables behave for neutron-rich systems.
Here we plan to extend the above study for the neutron-rich
systems and to look whether the above participant-spectator
demonstration still holds for systems lie far away from the
stability line. The present study is carried out within the
framework of isospin-dependent quantum molecular dynamics (IQMD)
model \cite{hart98}.
 \par
We simulate the reactions of Ca+Ca, Ni+Ni, Zr+Zr, Sn+Sn, and Xe+Xe
series having N/Z = 1.0, 1.6 and 2.0. In particular, we simulate
the reactions of $^{40}$Ca+$^{40}$Ca (105), $^{52}$Ca+$^{52}$Ca
(85), $^{60}$Ca+$^{60}$Ca (73); $^{58}$Ni+$^{58}$Ni (98),
$^{72}$Ni+$^{72}$Ni (82), $^{84}$Ni+$^{84}$Ni (72);
$^{81}$Zr+$^{81}$Zr (86), $^{104}$Zr+$^{104}$Zr (74),
$^{120}$Zr+$^{120}$Zr (67); $^{100}$Sn+$^{100}$Sn (82),
$^{129}$Sn+$^{129}$Sn (72), $^{150}$Sn+$^{150}$Sn (64) and
$^{110}$Xe+$^{110}$Xe (76), $^{140}$Xe+$^{140}$Xe (68) and
$^{162}$Xe+$^{162}$Xe (61) at an impact parameter of
b/b$_{\textrm{max}}$ = 0.2-0.4 at the incident energies equal to
 balance energy. The values in the brackets represent
the balance energies for the systems. We use a soft equation of
state along with the standard isospin- and energy-dependent cross
section reduced by
  20$\%$, i.e. $\sigma$ = 0.8 $\sigma_{nn}^{free}$.
The reactions are followed till the transverse in-plane flow
saturates. It is worth mentioning here that the saturation time
varies with the mass of the system. Saturation time is about 100
(150 fm/c) in lighter (heavy) colliding nuclei in the present
energy domain. We use the quantity "\textit{directed transverse
momentum $\langle p_{x}^{dir}\rangle$}" to define the nuclear
transverse in-plane flow, which is defined as \cite{sood1,hart98}

\begin {equation}
\langle{p_{x}^{dir}}\rangle = \frac{1} {A}\sum_{i=1}^{A}{sign\{
{y(i)}\} p_{x}(i)},
\end {equation}
where $y(i)$ and $p_{x}$(i) are, respectively, the rapidity
(calculated in the center of mass system) and the momentum of the
$i^{th}$ particle. The rapidity is defined as

\begin {equation}
Y(i)= \frac{1}{2}\ln\frac{{\vec{E}}(i)+{\vec{p}}_{z}(i)}
{{\vec{E}}(i)-{\vec{p}}_{z}(i)},
\end {equation}

where $\vec{E}(i)$ and $\vec{p_{z}}(i)$ are, respectively, the
energy and longitudinal momentum of the $i^{th}$ particle. In this
definition, all the rapidity bins are taken into account. It is
worth mentioning that the E$_{bal}$ has the same value for all
fragments types \cite{pak}. Further the apparatus corrections and
acceptance do not play any role in calculation of the E$_{bal}$.

\begin{figure}[!t] \centering
 \vskip -1cm
\includegraphics[width=14cm]{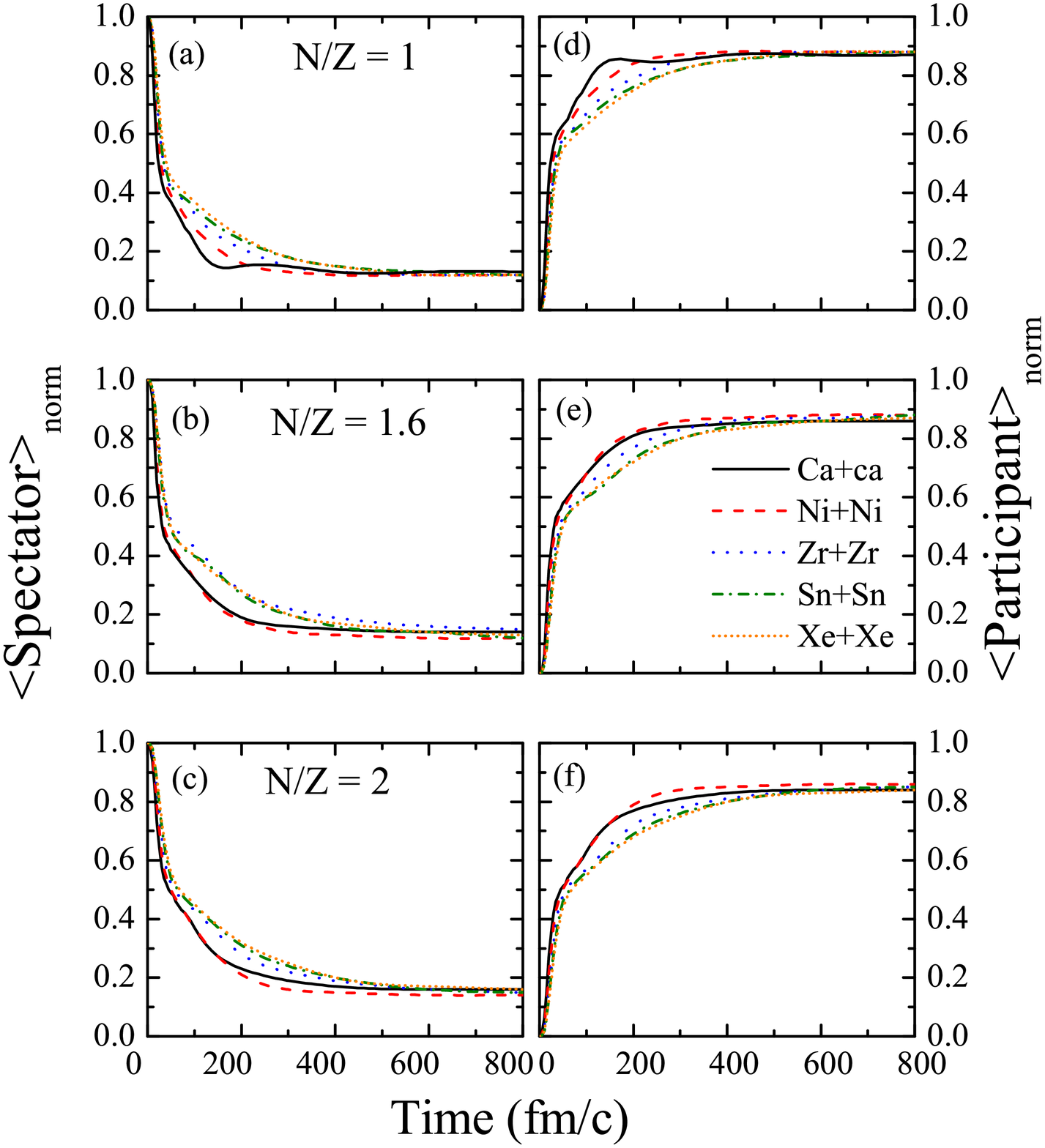}
\caption{(Color online) The time evolution of spectator matter
(left panels) and participant matter (right panels) for systems
having N/Z = 1.0, 1.6 and 2.0. Lines are explained in the
text.}\label{fig1}
\end{figure}

\begin{figure}[!t] \centering
 \vskip 1cm
\includegraphics[angle=0,width=10cm]{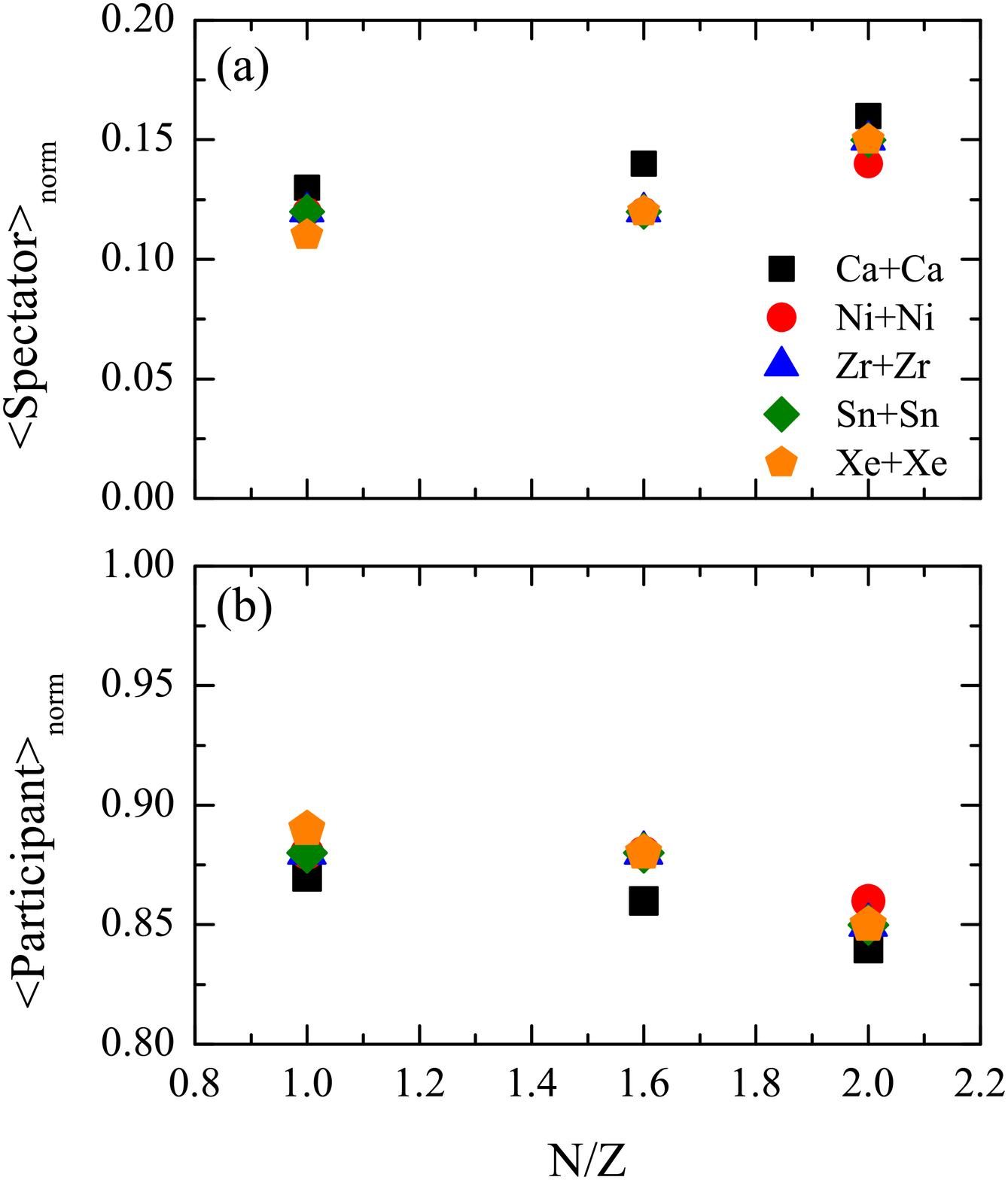}
 \vskip -0cm \caption{ (Color online) The N/Z dependence of participant and spectator matter. Symbols are explained in the text.} \label{fig2}
\end{figure}

\begin{figure}[!t] \centering
\vskip 0.5cm
\includegraphics[angle=0,width=10cm]{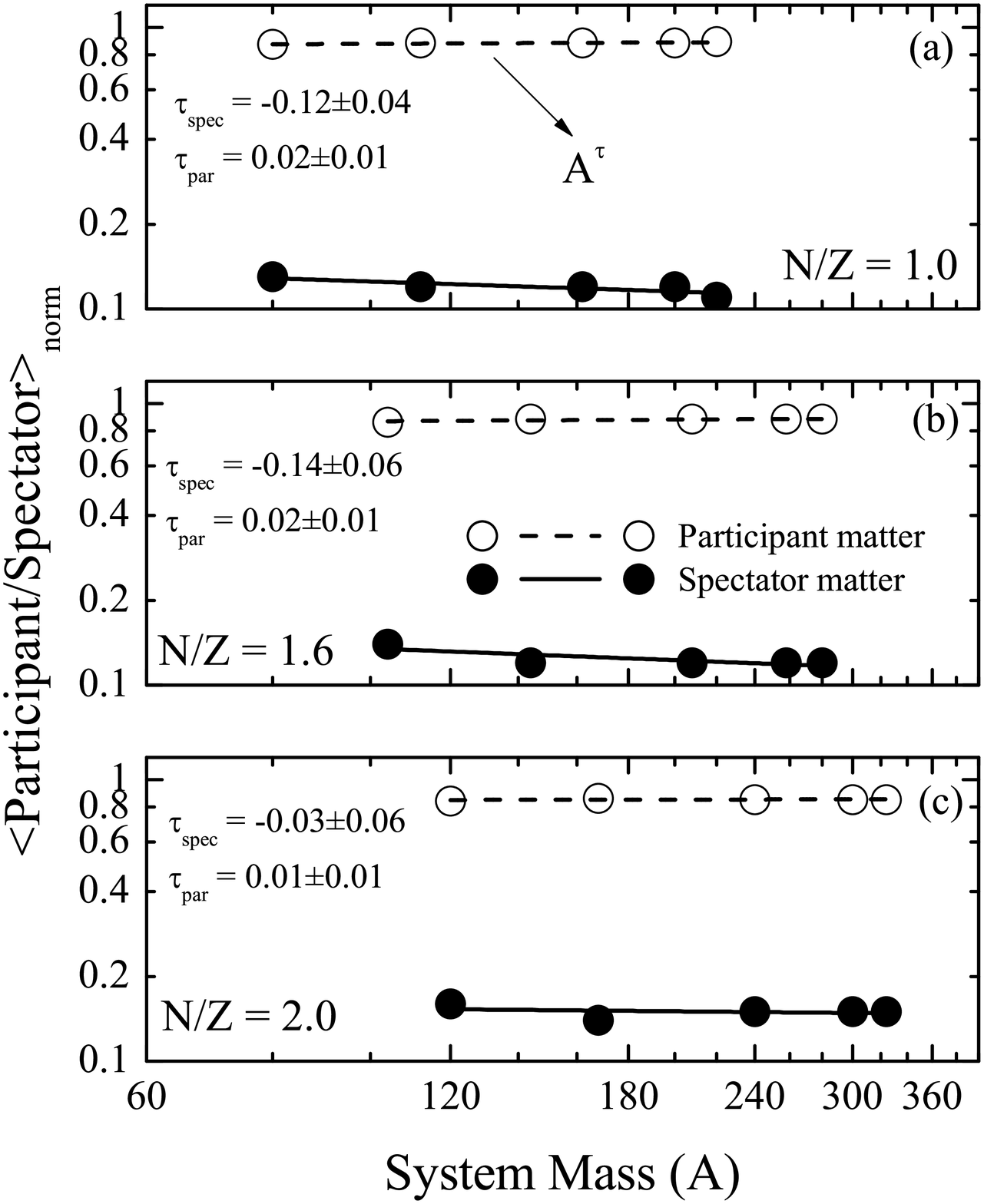}
\vskip 0.5cm \caption{(Color online) The system size dependence of
participant and spectator matter for different N/Z ratios. Various
symbols are explained in the text.}\label{fig3}
\end{figure}
\par
Since the balance energy represents the counterbalancing of
attractive mean filed potential and repulsive nucleon-nucleon
scattering and so this counterbalancing is reflected in
participant and spectator matter as predicted in Ref. \cite{sood1}
In the present study, we define the participant-spectator matter
in terms of nucleonic concept. All nucleons having experienced at
least one collision are counted as \emph{participant matter}. The
remaining matter is labeled as \emph{spectator matter}. These
definitions give us the possibility of analyzing the reaction in
terms of the participant-spectator fireball model.
\par
In fig. 1 we display the time evolution of normalized spectator
matter (left panel) and participant matter (right panel). The
upper, middle and lower panels represent the results for N/Z =
1.0, 1.6, and 2.0, respectively. Lines correspond to different
systems. Solid, dashed, dotted, dash-dotted, and short-dotted
lines represent the reactions of Ca+Ca, Ni+Ni, Zr+Zr, Sn+Sn, and
Xe+Xe, respectively. From figure, we find that at the start of the
reaction, all nucleons constitute the spectator matter. Therefore,
no participant matter exists at t = 0 fm/c. As the reaction
proceeds we have the decrease in spectator matter with
corresponding increase in participant matter. We also find that
for lighter systems like Ca+Ca and Ni+Ni, the transition from
spectator to participant matter is swift and sudden whereas for
the heavier colliding nuclei, the transition is slow and gradual
as predicted in Ref. \cite{sood1}. This is because of the fact
that lighter reactions occur at relatively high energies. At the
end of the reactions, we have nearly the same participant matter,
which indicates the universality in balancing the attractive and
repulsive forces. We also see that similar behaviour exists for
all N/Z ratios. This indicates that participant-spectator
behaviour is similar for neutron-rich systems as for systems lying
on the stability line (N/Z = 1).

\par
In fig. 2 we display the N/Z dependence of participant and
spectator matter. Upper (lower) panel displays the spectator
(participant) matter. Squares, circles, triangles, diamonds, and
pentagons represent the reactions of Ca+Ca, Ni+Ni, Zr+Zr, Sn+Sn,
and Xe+Xe, respectively. We find that for all the system masses
participant-spectator matter is almost independent of N/Z. There
is a very slight increase (decrease) in spectator (participant)
matter with N/Z of the system.

\par
In fig. 3, we display the system size dependence of the
participant and spectator matter. Open (solid) symbols represent
participant (spectator) matter. Upper, middle and lower panels
represent the results for N/Z = 1.0, 1.6 and 2.0, respectively. We
see that participant-spectator matter follows a power law
behaviour ($\propto$ A$^{\tau}$) with the system mass. The power
law factor is -0.12$\pm$ 0.04 (0.02$\pm$ 0.01), -0.14$\pm$ 0.06
(0.02$\pm$ 0.01), and -0.03$\pm$ 0.06 (0.01$\pm$ 0.01) for
spectator (participant) matter having N/Z = 1.0, 1.6 and 2.0,
respectively. Thus, a nearly mass independent behaviour is obeyed
by the participant and spectator matter for all the N/Z ratios.


\begin{figure}[!t] \centering
 \vskip -1cm
\includegraphics[width=10cm]{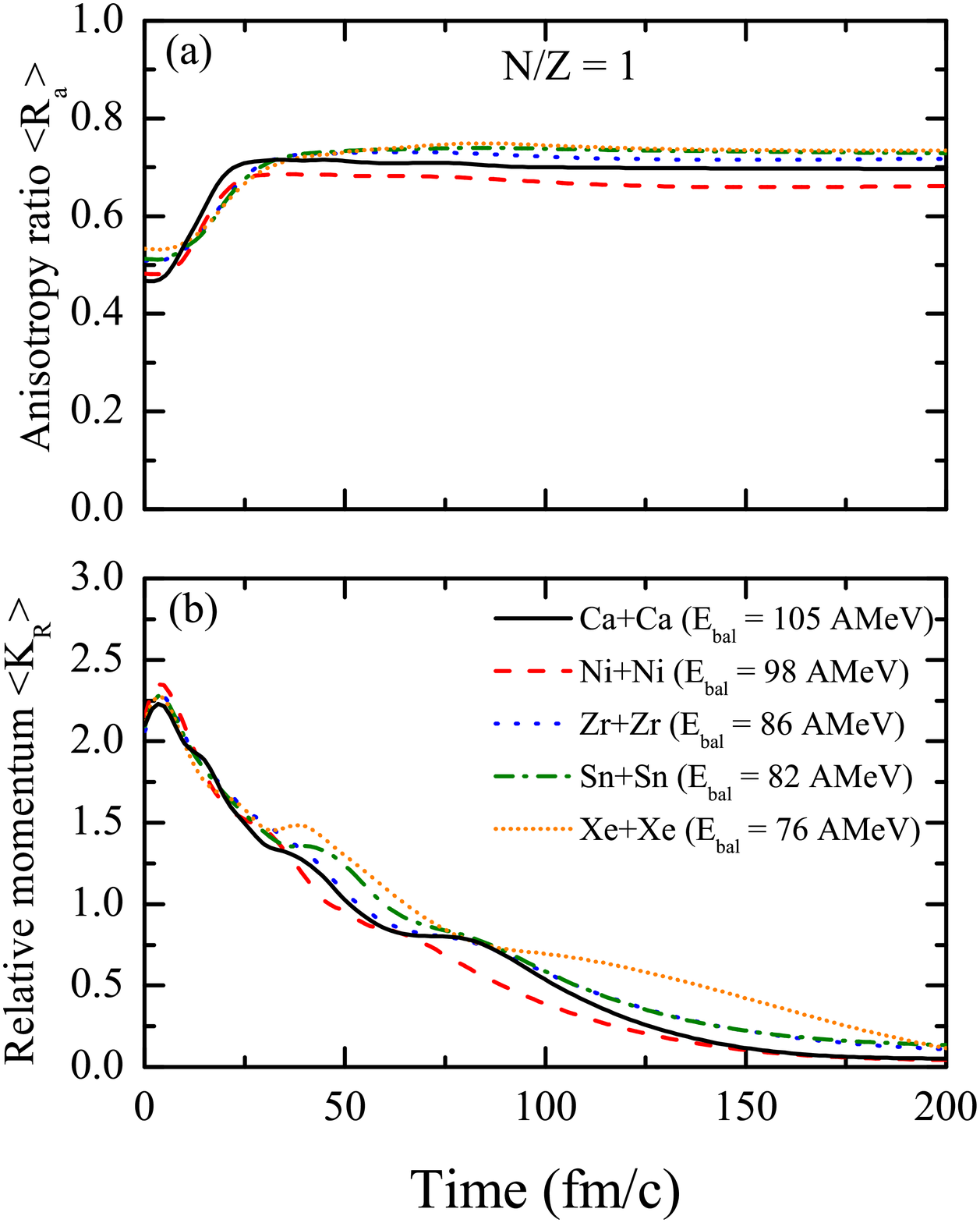}
\caption{(Color online) The time evolution of the anisotropy ratio
(upper panel) and relative momentum (lower panel) for various
systems having N/Z = 1.0. Lines have the same meaning as in Fig.
1}\label{fig4}
\end{figure}
\par
In fig. 4 we display the time evolution of anisotropy ratio
$<R_{a}>$ (upper panel) and relative momentum $<K_{R}>$ (lower
panel) for different system masses having N/Z = 1.0. The $<R_{a}>$
is defined as
\begin{equation}
<R_{a}> =
\frac{\sqrt{p_{x}^{2}}+\sqrt{p_{y}^{2}}}{2\sqrt{p_{z}^{2}}}.
\end{equation}

This anisotropy ratio is an indicator of the global equilibrium of
the system. This represents the equilibrium of the whole system
and does not depend on the local positions. The full global
equilibrium averaged over large number of events will correspond
to $<R_{a}>$ = 1. The second quantity, the relative momentum
$<K_{R}>$ of two colliding Fermi spheres, is defined as

\begin{equation}
<K_{R}> = <|\vec{P}_{P}(\vec{r},t)-\vec{P}_{T}(\vec{r},t)|/\hbar>,
 \label{kr}
\end{equation}
 \begin{figure}[!t] \centering
 \vskip -1cm
\includegraphics[width=10cm]{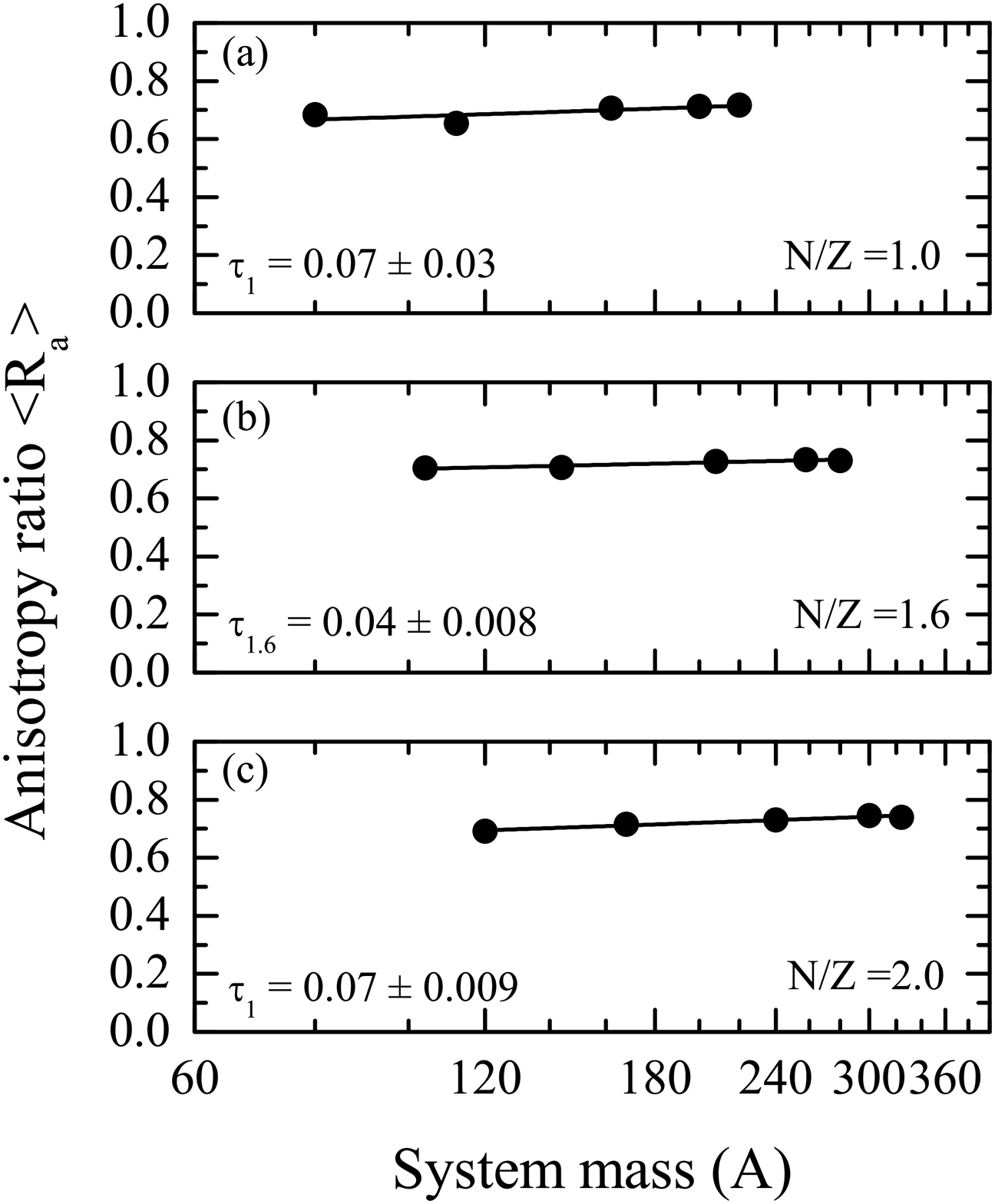}
\caption{(Color online) The system size dependence of anisotropy
ratio for various N/Z ratios.}\label{fig5}
\end{figure}


where

\begin{equation}
\vec{P_{i}}(\vec{r},t) =
\frac{\sum_{j=1}^{A}\vec{P_{j}}(t)\rho_{j}(\vec{r},t)}
{\rho_{j}(\vec{r},t)}~~~~~         i=1,2.
\end{equation}
Here $\vec{P_{j}}$ and $\rho_{j}$ are the momentum and density of
the \emph{j}th particle and \emph{i} stands for either projectile
or target. The $<$$K_{R}$$>$ is an indicator of the local
equilibrium because it depends also on the local position
\emph{r}.

From figure 4(a) (upper panel), we see that anisotropy ratio
increases as the reaction proceeds and finally saturates after the
high density phase is over. We also see that the influence of
system size is very less on anisotropy ratio and hence indicates
towards the equilibrium of the system. From fig. 4(b) (lower
panel) we see that relative momentum decreases as the reaction
proceeds and the smaller value of $<$$K_{R}$$>$ at the end of the
reaction indicates toward the better thermalization of the matter.
 We also see from the figure that $<$$R_{a}$$>$
ratio saturates as soon as high density phase is over which
signifies that the nucleon-nucleon collisions happening after high
density phase do not change the momentum space significantly.
\par
In fig. 5 we display the system size dependence of anisotropy
ratio for systems having N/Z ratios 1.0, 1.6 and 2.0. From figure
we see that anisotropy ratio follows a power law behaviour
($\propto$ A$^{\tau}$) with system size. The power law factor is
0.07 $\pm$0.03, 0.04 $\pm$0.008, and 0.07 $\pm$0.009 for N/Z
ratios 1.0, 1.6 and 2.0, respectively.
\par
In summary, we studied the participant-spectator matter at the
energy of vanishing flow for neutron-rich systems. The study
revealed a similar behaviour of participant-spectator for
neutron-rich systems as for stable systems and also pointed
towards a nearly mass dependence behaviour of
participant-spectator matter of neutron-rich systems at the energy
of vanishing flow. Similar mass independent behaviour is also
found for the anisotropy ratio.
 \par
This work has been supported by a grant from Centre of Scientific
and Industrial Research (CSIR), Govt. of India and Indo-French
Centre For The Promotion Of Advanced Research (IFCPAR) under
project no. 4104-1.

\end{document}